\documentclass{vienna-conf2019}
\usepackage{graphicx}
\usepackage{hyperref}
\usepackage[]{natbib}  
\usepackage{epstopdf}

\def\BibTeX{{\rm B\kern-.05em{\sc i\kern-.025em b}\kern-.08em
    T\kern-.1667em\lower.7ex\hbox{E}\kern-.125emX}}
\bibpunct{(}{)}{;}{a}{}{,}  

\begin{document}

\TitreGlobal{Stars and their variability observed from space}


\title{Shine BRITE: \\
shedding light on stellar variability through advanced models}

\runningtitle{Advanced atmospheric modeling for understanding the variability of stars}

\author{D. Fabbian}\address{Georg-August-Universit{\"a}t G{\"o}ttingen, Institut f{\"u}r Astrophysik, Friedrich-Hund-Platz 1, D-37077 G{\"o}ttingen, Germany}

\author{F. Kupka}\address{MPI f{\"u}r Sonnensystemforschung, Justus-von-Liebig-Weg 3, D-37077 G{\"o}ttingen, Germany}

\author{D. Kr{\"u}ger$^{1,2,}$}\address{Universit{\"a}t Wien, Fakult{\"a}t f{\"u}r Mathematik, Oskar-Morgenstern-Platz 1, A-1090 Wien, Austria}

\author{N.\,M. Kostogryz$^{2}$}

\author{N. Piskunov}\address{Uppsala University, Department of Physics and Astronomy, Regementsv{\"a}gen 1, SE-752 37 Uppsala, Sweden}

\setcounter{page}{237}


\maketitle


\begin{abstract}
The correct interpretation of the large amount of complex data from next-generation (in particular, space-based) observational facilities requires a very strong theoretical underpinning. 
One can thus predict that in the near future the use of atmospheric models obtained through the use of three-dimensional (3-D) radiation magnetohydrodynamics (RMHD) codes, coupled with advanced radiative transfer treatment including non-local thermodynamic equilibrium (non-LTE) effects and polarisation, will become the norm. 
In particular, stellar brightness variability in cool (i.~e., spectral type F, G, K, and M) stars can be induced by several different effects, besides pulsation. We briefly discuss some literature results and mention some of our recent progress. Finally, we attempt to peek into the future of understanding this important aspect of the life of stars.
\end{abstract}

\begin{keywords}
Magnetohydrodynamics (MHD), radiative transfer, Stars: variables: general
\end{keywords}


\section{Introduction}

Numerical radiation (magneto-)hydrodynamic simulations of stellar atmospheres provide a strong basis for a multitude of studies on different aspects of, in particular, cool stars. One of the topics of great interest is understanding why many -- if not all -- of them undergo some amount of variation in their brightness with time. Due to its proximity, one particularly well-studied case is the solar one. However, because the Sun's periodic brightness variation is small (about $0.1\%$) and has a relatively long period of 11 years, it was discovered relatively recently compared to the variations observed for other stars (intrinsically-variable ones such as pulsating or eruptive or cataclysmic/explosive variables, or extrinsic-variable ones such as eclipsing or rotating variables), which have been known in some cases for already many centuries. 

Below, we provide a very brief discussion of some recent results of interest, followed by concise conclusions. 
 
\section{Discussion}

``Box in a star'' numerical simulations of stellar atmospheres, based on solving the (numerical) radiation (magneto-)hydrodynamics equations, aim at reproducing and understanding features of (magneto-)convection as observed by ground- and space-based instruments. 

These type of simulations are well-tested, being capable of reproducing observations, e.g. of solar granulation and stellar spectral lines. They are, moreover, useful to gain physical insight into convective heat transport and on the interaction of convection with pulsation modes. 

\begin{figure}[ht!]
 \centering
 \includegraphics[width=0.99\textwidth,clip]{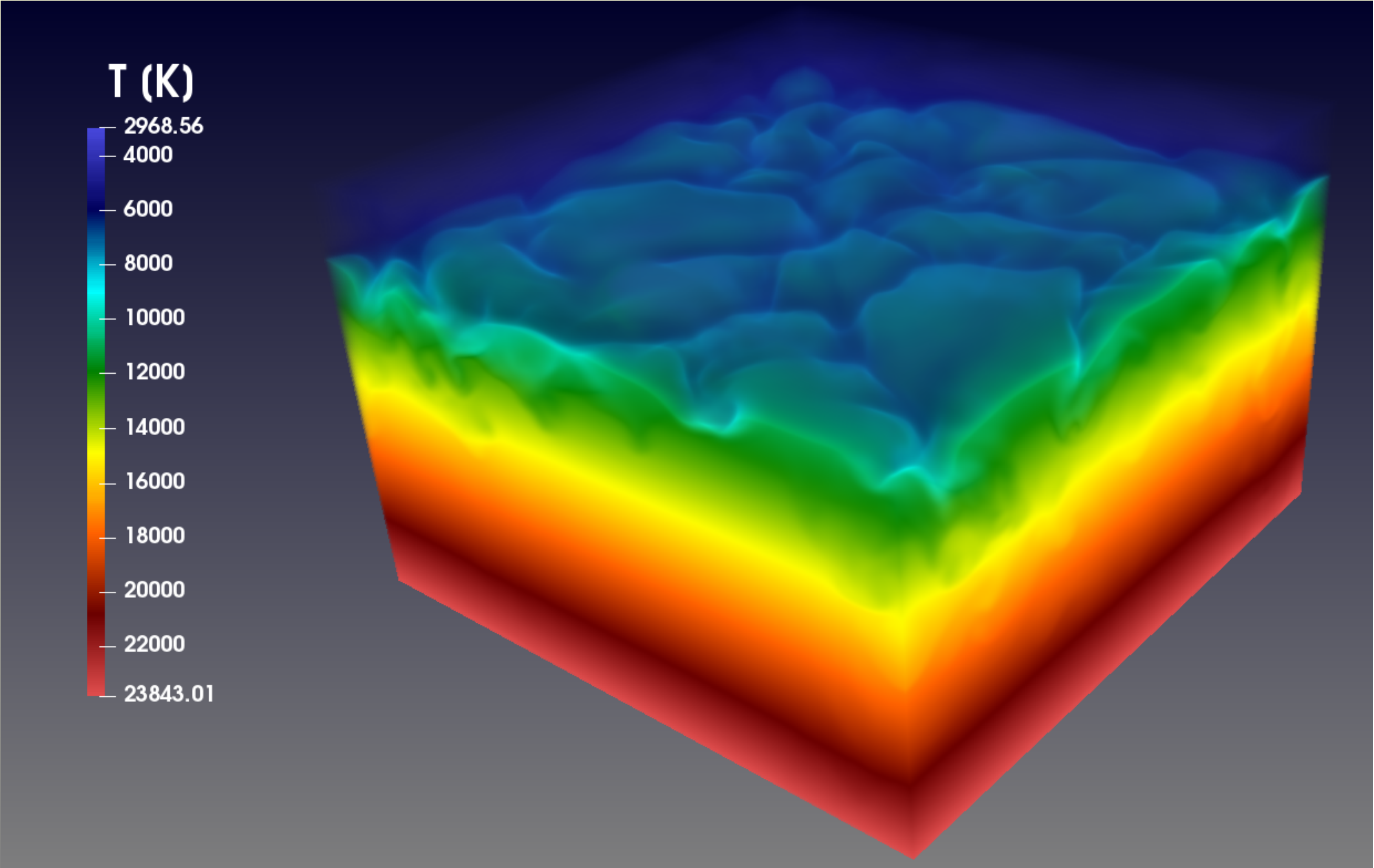}
  \caption{Temperature (in K units) for a snapshot 
of the solar-like simulation ``SLOPMD1'' performed with the code ANTARES and visualised using ParaView (\url{https://www.paraview.org/}). This is the snapshot used as input to obtain the synthetic spectrum shown in Fig.\,\ref{fabbian:fig3}.}
  \label{fabbian:fig1}
\end{figure}

In Fig.\,\ref{fabbian:fig1} we show the temperature for a snapshot from the $\sim21$~solar h of statistically-stable evolution of a solar-like simulation of $\sim 3.7$~Mm vertical extent and of $6.0$~Mm extent in each horizontal direction. 

Improvements in the radiative transfer treatment and in the input starting modeling as well as the possibility of using, where appropriate, the (non-grey, 3-D) Eddington approximation for faster calculations have recently been implemented in the code (see, e.~g., \cite{2019JPhCS1225a2017K, IAU2018, 2019SoPh..???..???K}). 

ANTARES has recently been successfully used by our group to perform long-duration 3-D simulations of solar and stellar convection to study p-mode excitation and damping processes. The spectral power of vertical velocity in the solar granulation simulation performed with this code shows eigenmode frequencies with the correct shape. Scaling has to be carried out to account for the limited size and depth coverage of the simulated model (i.~e., for the relative shallowness of the ``box in a star'' geometry compared to the whole star). 
Apart from the needed scaling, it is important (as presented here by F. Kupka, keynote talk 5k02) that the 3-D hydrodynamical simulations have sufficient spatial and temporal resolution, spatial vertical and horizontal extent, and time duration in order to successfully be applied for asteroseismology studies such as excitation and damping of solar-like p-modes. In particular, as is the case for our simulations, that they are able to well describe the properties of the superadiabatic layer, one of the crucial aspects in order to match the observed p-mode characteristics (frequency and amplitude and shape of lines) in the power spectra of solar-like oscillating stars. It will of course be interesting to see also MHD models applied for this purpose in the future, to understand the influence of magnetism on stellar atmospheric plasma, as we will briefly discuss below. 

Brightness variations observed in the light curve of different stars can, however, at least in part, be caused by processes other than the solar-like oscillations (i.~e., those excited by convective motions). 

Granulation is the clearest visual manifestation at the surface of stars of the process of convection. Due to the stochastic nature of the latter, the granulation pattern produces micro-variations in time of the emerging radiation. Once p-mode oscillations (manifesting as discrete peaks) are removed from the solar power spectrum, for example, it is mostly this granulation signal (so-called, ``granulation background'') which is left. 

One sub-category of cool stars is particularly interesting, namely the one composed of stars on the main sequence and broadly similar to the Sun (spectral types F, G, and K), so-called solar-type stars. The variability of their magnetic activity has recently been reviewed e.~g. in \cite{2017AN....338..753F}. 

\begin{figure}[ht!]
 \centering
 \includegraphics[width=0.75\textwidth,clip]{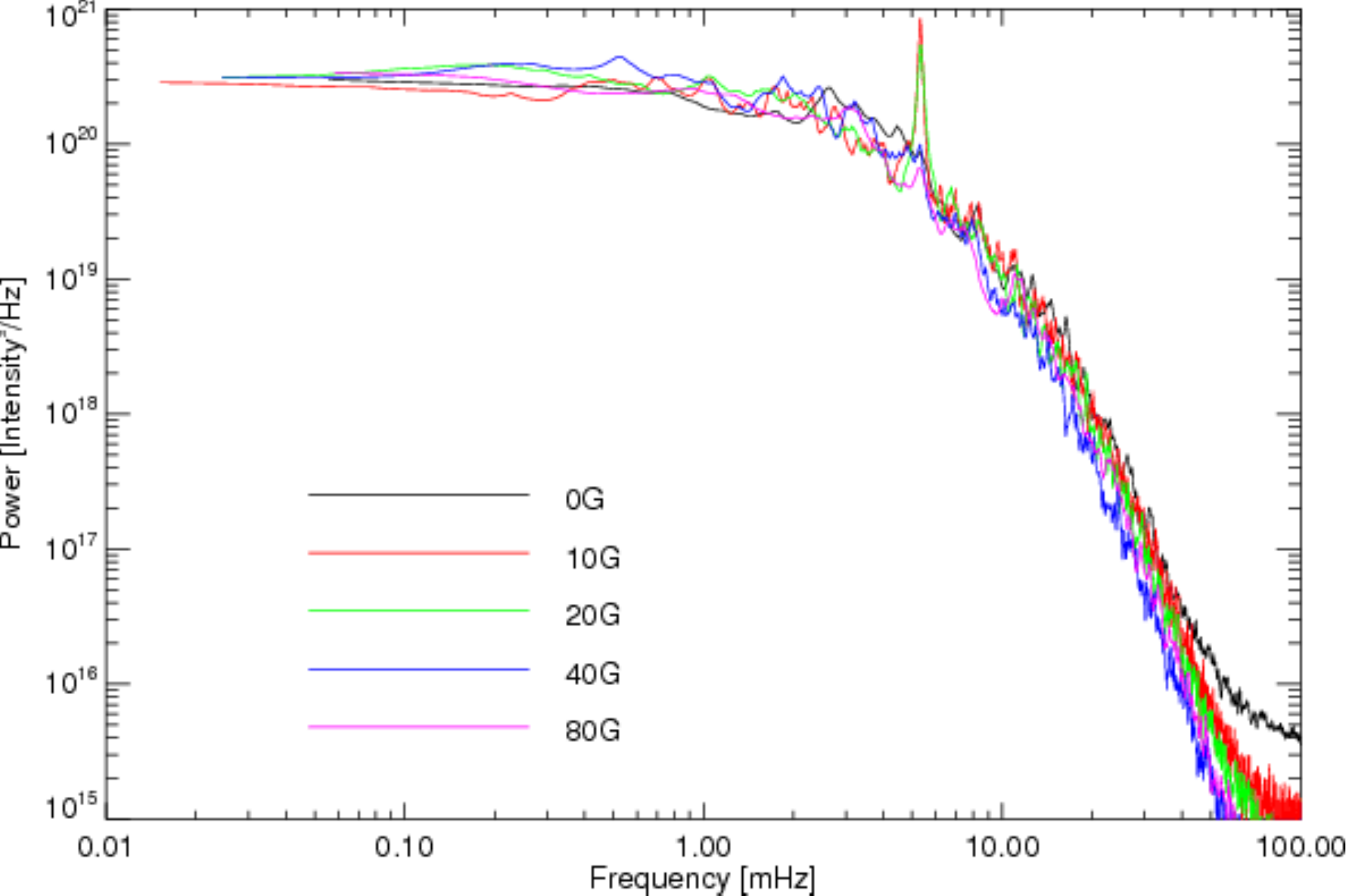}
  \caption{Temporal power spectra of the horizontally-averaged vertically-emergent intensity of different MHD runs (solid curves of different colors) with solar atmospheric parameters (image adapted from Fig. 5 of \cite{2009A&A...506..167L}).}
  \label{fabbian:fig2}
\end{figure}

For the Sun, observations show a steeply-decreasing granular signal power at high 
($\sim 1-8$~mHz) frequencies, while the roughly constant power around $\sim 0.2$~mHz 
is followed by an increase with decreasing frequencies below $\sim 0.1$~mHz caused by 
solar magnetic activity. 
Using the CO$^{5}$BOLD code, \cite{2009A&A...506..167L} showed that atmospheric models 
from 3-D radiation hydrodynamic (RHD) simulations, while able to reproduce the 
granulation-related brightness fluctuations power spectrum at intermediate and 
high frequency reasonably well, contain -- as to be expected, being representative 
of the pure (non-MHD) case -- less power at low frequencies than in the observational 
data. 
They found a puzzling discrepancy with the observed granulation background for F-type 
dwarfs, showing that only after scaling the power and frequency of their
predicted granulation-related brigthness fluctuations to account e.~g. for possible
uncertainties in stellar parameter determinations and after adding an ad-hoc,
power-law magnetic activity signal, the mismatch could at least to a significant
extent be resolved.
The authors then further performed 2-D RMHD solar simulations assuming different levels 
of initial magnetic field strength.
They found that observations have a low level of granulation-related brightness 
fluctuations around $1$~mHz compared to their 2-D RMHD predictions. 
Thanks to the very similar topology of granular flows among 
solar-type stars, they make the hypothesis that the rise at low frequency in the 
temporal power spectrum of F-type dwarfs, may be an indication of some type of magnetic 
activity also for those stars, in analogy to what hinted in the case of observational 
data for the Sun. 
Their 2-D results, however (see their derived temporal power spectra of resulting 
emergent intensity in Fig.\,\ref{fabbian:fig2}), exclude local dynamo action in the 
granular flow as the source of magnetism producing the mismatch between predictions 
and observations. This leaves the existence of a ubiquitous, larger-scale magnetic field of several hundred G (not yet 
detected by observations, however) as the only plausible cause able to affect the dynamics of granulation to an extent of the order of that needed to reduce the brightness 
fluctuations predictions to the observed low level.

It remains to be seen if switching to RMHD simulations in 3-D, 
and targeted to solar-type stars having very precise stellar parameter determinations, 
can clarify the appearance of the 
brightness fluctuations power spectra for solar-type stars other than the Sun, or if 
observations will confirm the presence of strong magnetic field organised on larger 
scale, or if yet other explanations may be required. 

\begin{figure}[ht!]
 \centering
  \includegraphics[width=0.99\textwidth,clip]{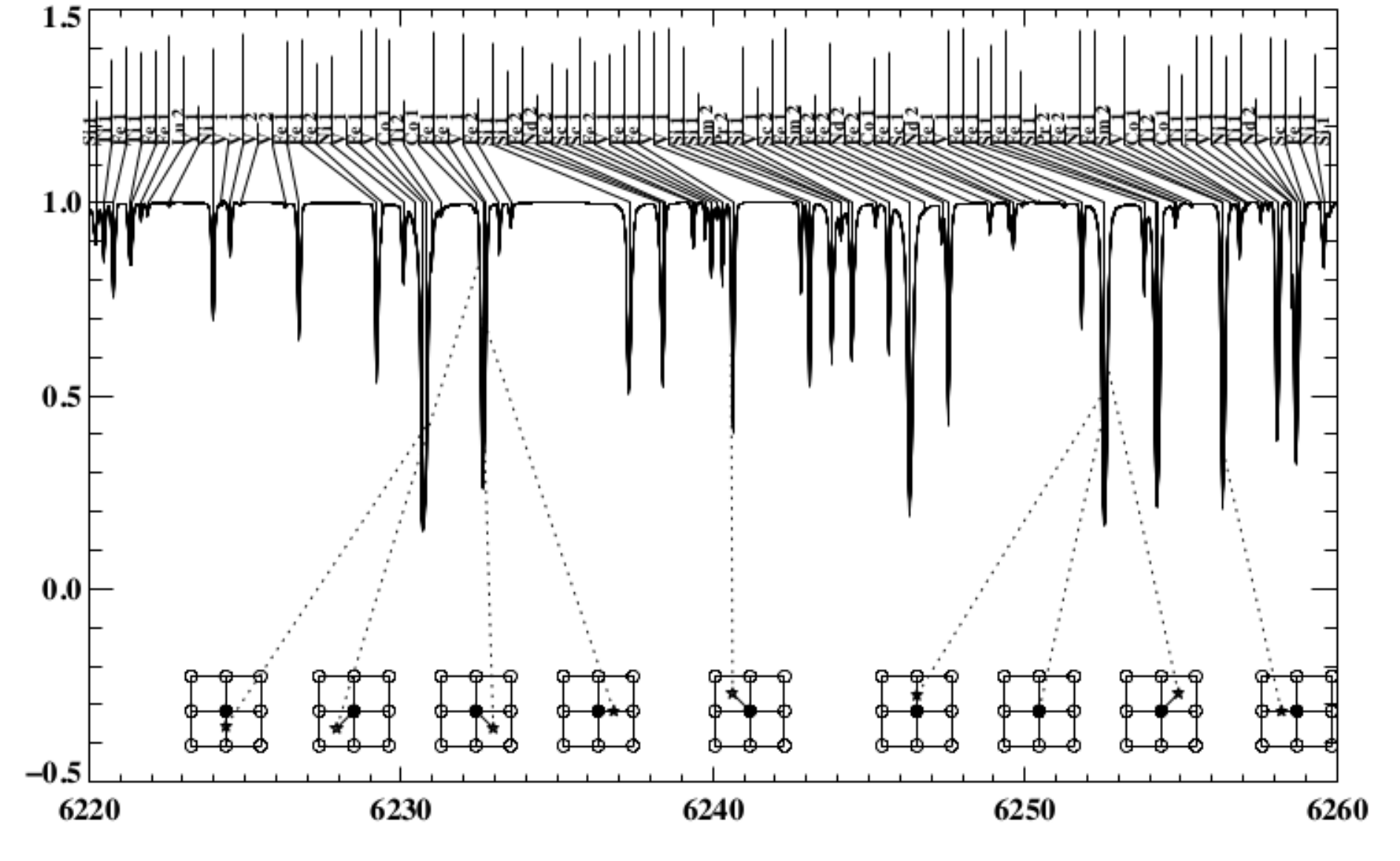}
  \caption{Disk-center intensity for the synthetic spectrum obtained for the region $622-626$~nm based on a 3-D RHD solar simulation performed with the ANTARES code. The spectrum was calculated using as input the snapshot shown in Fig.\,\ref{fabbian:fig1}.}
  \label{fabbian:fig3}
\end{figure}

Spectral line synthesis tools allow one to compare with and check against observations the synthetic intensity spectrum obtained on the basis of the multi-dimensional stellar photosphere simulations. We give here (Fig.\,\ref{fabbian:fig3}) an example synthetic spectrum computed on the basis of a solar granulation model from an ANTARES radiation hydrodynamic photospheric simulation having ``Model S'' \cite{1996Sci...272.1286C} as starting input model and then relaxed as described in \cite{2017LRCA....3....1K}.

\cite{2010ApJ...724.1536F, 2012A&A...548A..35F, 2015ApJ...802...96F} showed that the presence of magnetic fields can have a significant effect on the formation of Fe and O spectral lines in 3-D MHD solar simulations, affecting their profile and intensity and thus the photospheric chemical abundances inferred from a fit to observations. 

Some spectral lines have particular sensitivity to the level of solar and stellar activity (see, e.~g., \cite{2009A&A...499..301V}).

Concerning the Sun, very recently \cite{2019SoPh..???..???C} performed a thorough study comparing the results of different commonly-employed radiative transfer codes. They showed the importance of being aware of the relevant techniques, approximations and uncertainties involved, to explain the subtle differences they found between the synthetic quiet-Sun spectra and thus to well reproduce solar irradiance measurements and understand which features contribute the most to solar variability. 

Of particular usefulness is the effort on the side of theoretical and experimental physicists to achieve more accurate atomic and molecular data. Databases listing the most updated such data and compilations, e.g. the VALD3 (the current version of the VALD online database, as described in \cite{2017ASPC..510..518P}) are of great help to improve the atmospheric and spectral synthesis modeling and should be strongly supported by the stellar astrophysical community. 

\section{Conclusions}

We briefly discussed some issues related to the understanding of stellar brightness variability. Cool stars show variability with periods ranging from very short to very long timescales, with variations of different type often overlapping. Thus, the physical causes of these different variations are not easy to disentangle. One main point is that, being based on first-principles and updated micro- and macro-physics and being able to withstand demanding tests against observations, 3-D RMHD stellar atmospheric models are nowadays sufficiently well-developed to be employed also in this field as the required input. 

Naturally, this hints at the idea that the application of advanced stellar atmospheric simulations including ever-more-complete treatment of relevant physical processes is one of the main avenues for better understanding also this aspect of the life of stars and will become more widespread and common practice in place of outdated and oversimplified (e.g., one-dimensional, plane-parallel, static, gray, scattering- and magnetic-free atmosphere) modeling.

\begin{acknowledgements}
DF, FK, and DK gratefully acknowledge support by the Austrian Science Fund (FWF) through project P29172-N27, and they also thank the Vienna Scientific Cluster VSC-3 (project 70950) and the Faculty of Mathematics at the University of Vienna for the granted computing resources. FK acknowledges partial support through the European Research Council Synergy Grant WHOLE SUN \#810218. NMK is supported by the German space agency (Deutsches Zentrum f{\"u}r Luft- und Raumfahrt) under PLATO Data Center grant 50OO1501. 
\end{acknowledgements}

\bibliographystyle{aa}  
\bibliography{fabbian_4o01} 

\end{document}